\documentclass[a4paper]{jpconf}

\usepackage{amsmath,amstext}
\usepackage{apjfonts} 
\usepackage{graphicx}
\usepackage[square,sort&compress]{natbib}
\usepackage{appendix}

\usepackage{color}
\usepackage{xcolor}

\begin{document}

\title{Coupled MHD -- Hybrid Simulations of Space Plasmas }

\author{S.~P. Moschou$^1$,
I.~V. Sokolov$^2$,
O. Cohen$^3$,
G. Toth$^2$,
J.~J. Drake$^1$,
Z. Huang$^2$,
C. Garraffo$^1$,
J.~D. Alvarado-G\'omez$^1$, 
T. Gombosi$^2$}

\address{$^1$ Center for Astrophysics {$\vert$} Harvard \& Smithsonian, 60 Garden Street, Cambridge, MA 02138, USA}
\address{$^2$ Center for Space Environment Modeling, University of Michigan, Ann Arbor, MI 48109, USA}
\address{$^3$ Lowell Center for Space Science and Technology, University of Massachusetts, Lowell, MA 01854, USA}

 \eads{\mailto{sofia.moschou@cfa.harvard.edu}}

\begin{abstract}
Heliospheric plasmas require multi-scale and multi-physics considerations.
On one hand, MHD codes are widely used for global simulations of the solar-terrestrial environments, but do not provide the most elaborate physical description of space plasmas. Hybrid codes, on the other hand, capture important physical processes, such as electric currents and effects of finite Larmor radius, but they can be used locally only, since the limitations in available computational resources do not allow for their use throughout a global computational domain. In the present work, we present a new coupled scheme which allows to switch blocks in the block-adaptive grids from fluid MHD to hybrid simulations, without modifying the self-consistent computation of the electromagnetic fields acting on fluids (in MHD simulation) or charged ion macroparticles (in hybrid simulation). In this way, the hybrid scheme can refine the description in specified regions of interest without compromising the efficiency of the global MHD code.
\end{abstract}

\section{Introduction}

The nature of solar eruptive events is still an open field of research. Currently, our understanding of energy partition in solar eruptive events is rather poor, due to the fact that both fluid and particle scales are involved. This problem requires sophisticated, and often times computationally expensive modeling. Currently missing from the modeling of solar eruptive events is a unified approach that captures both small and large scale dynamics involved in the production of Solar Energetic Particles (SEPs), flares and Coronal Mass Ejections (CMEs) \citep{Fletcher.etal:11}. Including a description of physics on both particle scales and scales of the order of solar radius is a very challenging and computationally expensive task. One way to tackle this problem is to couple MagnetoHydroDynamic (MHD) and hybrid models. 

In most hybrid codes, the ions are modeled as particles using Particle-In-Cell (PIC) methods, and the electrons are modeled as a fluid. Fluid codes have demonstrated great success in modeling very different astrophysical systems, especially on large scales, when the particle gyroradius can be considered infinitely small. However, when physical processes in {\color{black} scales} comparable to the particle gyroradius {\color{black} or plasma skin depth} cannot be neglected, particle methods are {\color{black} used}. 
In the course of {\color{black}historical} evolution, modern versions of the PIC method \citep{Birdsall05}, on one hand, and high-resolution conservative numerical schemes for Computational Fluid Dynamics (CFD) \citep{Hirsch97}, on the other, have diverged drastically. However, there was a time when they were {\color{black}close} conceptually and some schemes were proposed allowing natural combination of their advantages. The PIC method was originally proposed by \cite{Harlow:64} for solving CFD problems and was already successfully applied on complicated dynamical problems in the 60s. At that time, however, the flaws of this {\color{black}method} were also revealed. For example, spurious oscillations of the flow functions were appearing for simulations with a low number of particles, which could be mitigated by increasing the number of particles. 

In this work, we use the experience gained for decades in hybrid code development, to implement a hybrid numerical scheme within the Space Weather Modeling Framework (SWMF) by \cite{Toth.etal:12} and couple it with the MHD fluid model Block-Adaptive-Tree-Solarwind-Roe-Upwind-Scheme (BATS-R-US) by \cite{powell99}. The modular nature and advanced software engineering of the SWMF makes one- and two-way coupling between individual models straightforward for the user. The end goal of this work is to simulate large-scale plasma systems, i.e., the solar corona, using coupled MHD--hybrid setups. This work is inspired by magnetospheric work of \cite{Daldorff.etal:14} and \cite{Toth.etal:16}, who developed a new modeling capability called MHD--EPIC, to embed the implicit PIC model iPIC3D into localised regions of the global BATS-R-US domain. Similar to the role that the iPIC3D model plays for the MHD--EPIC scheme of \cite{Toth.etal:16}, the hybrid algorithm will provide a more robust particle description in regions of interest, such as reconnection regions. The MHD--EPIC scheme, while very efficient for magnetospheric simulations of smaller bodies, such as planets and their satellites \cite[see][]{Chen.etal:17}, even with the artificial scaling of the ion inertial length scale to match the system's global MHD scales, is still too expensive for modeling the solar corona. The MHD--hybrid scheme provides a viable alternative in the spirit of MHD--EPIC for simulating the solar corona efficiently, as it is arguably computationally cheaper than simulating the full corona with purely PIC or even coupled MHD--EPIC models.

We have chosen to couple MHD and hybrid models that share the same two-step staging. We first start by explaining to the reader the advantages that such a choice may have and then proceed to the numerical schemes and methods used for the hybrid code.
In Section ~\ref{sec:fluid}, we describe similarities in the numerical schemes of both fluid and particle schemes. Subsection ~\ref{sec:hd} deals with cases where the dominant force is the pressure gradient, while cases where the Lorentz force is the dominant one are treated in Subsection ~\ref{sec:mhd}.
We then proceed to Section ~\ref{sec:hybrid} where we describe the hybrid algorithm step-by-step, exploiting the parallels drawn in Section ~\ref{sec:fluid}.
Finally, we end this paper with our conclusions in Section ~\ref{sec:discussion}.

\section{Parallels Between Fluid and Particle Schemes}\label{sec:fluid}
When combining physically different models within a common domain, it is preferable to choose as similar numerical methods as possible, using the same or compatible staging, conservation properties, etc, in order to facilitate the coupled simulation. 
This justifies our choice to combine particles and fluids within a single code. In the following section we elaborate on the parallels between the two methods.

\subsection{Coarse Particles \textit{versus} Fluid Particles}\label{sec:hd}

\begin{figure*}[htbp]
 \begin{center}
 \includegraphics[width=\textwidth]{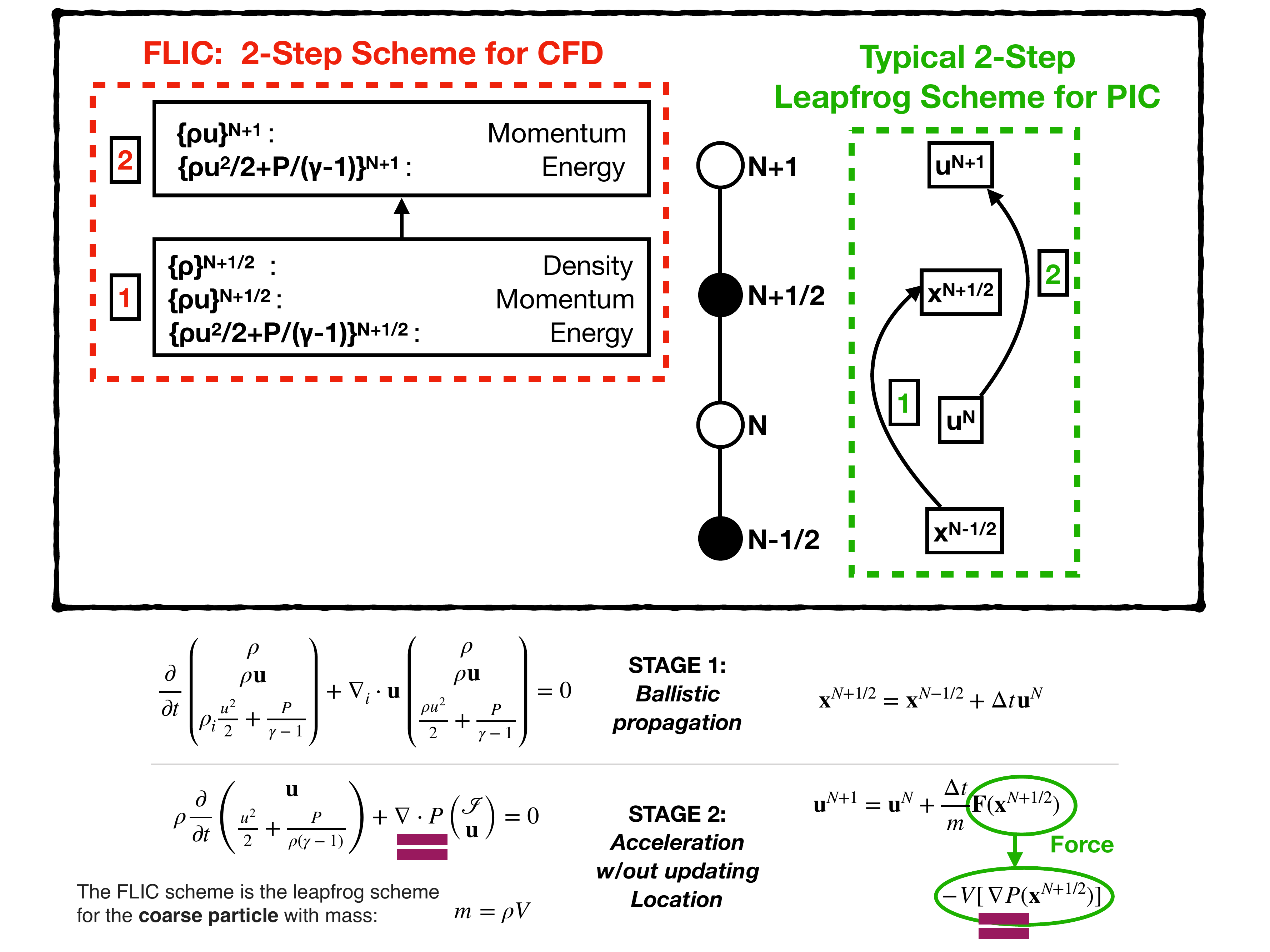}  \caption{Schematic summarizing the parallels in the staging between coarse particles and fluid particles leading to an enhanced efficiency.}
  \label{fig:flic1}
  \end{center}
  \end{figure*}

The high computational requirements of particle simulations lead to modifications of the PIC algorithm, designed to reduce the computational cost while still exploiting its advantages. The modified method named the Fluid-in-Cell, or FLIC, was first presented by \cite{Gentry66}. 
Both the FLIC method and the original PIC scheme \citep{Harlow:64} treat (1) the effect of particle (or fluid element) motion (the ``Lagrangian stage'') and (2) the particle acceleration as the result of interaction with the force field discretized at some Eulerian grid (the ``Eulerian stage'') as two sequential steps of the algorithm, as shown in Fig.~\ref{fig:flic1}.  In the FLIC method, both the Euler (particle) and Lagrangian steps are conservative.

\textbf{Stage 1:} Within the framework of the FLIC scheme for CFD, at the Lagrangian step, the mass, momentum and energy conservative transport is performed. Following the finite volume methodology, the densities of fluid mass, $\rho$, momentum, $\rho\mathbf{u}$, and energy, $\rho \frac{u^2}2+\frac{P}{\gamma-1}$, where $\mathbf{u}, P, \gamma$ are the velocity vector, pressure, and the adiabatic index respectively,
are updated as follows:
\begin{equation}\label{eq:LagrangianStep}
\frac{\partial}{\partial t}\left(
\begin{array}{c}\rho\\\rho\mathbf{u}\\ \rho_i\frac{u^2}2+\frac{P}{\gamma-1}\end{array}\right)_i+\nabla_i\cdot\mathbf{u}\left(
\begin{array}{c}\rho\\\rho\mathbf{u}\\ \frac{\rho u^2}2+\frac{P}{\gamma-1}\end{array}\right)=0.
\end{equation}
Herewith, the particle derivative in time means for the mass density:$\frac{\partial \rho_i}{\partial t}=\frac{\rho_i(t+\Delta t)-\rho_i(t)}{\Delta t},$  
and analogously for other state variables, assigned to given $i$-th control volume, $V_i$. The conservative divergence operator in Eq.~(\ref{eq:LagrangianStep})
is expressed via the fluxes \cite[see][]{Hirsch97} across the faces between this volume and neighboring, control volumes enumerated with the index, $i^\prime$. For the mass flux one has: $\nabla_i\cdot\rho\mathbf{u}=\frac1{V_i}\sum_{i^\prime}\mathbf{S}_{ii^\prime}\cdot\left(\rho\mathbf{u}\right)_{ii^\prime}$, with the face area vectors, $\mathbf{S}_{ii^\prime}$, being directed outward from the control volume in hand, and the {\it numerical flux}, $\left(\rho\mathbf{u}\right)_{ii^\prime}$, being expressed in terms of $\rho_i,\rho_{i^\prime},\mathbf{u}_i,\mathbf{u}_{i^\prime}$ \cite[see][]{Gentry66,Grigoryev:12}. 
This approach ensures the conservation of total mass, $M=\sum_i{V_i\rho_i}$: 
$$\frac{dM}{dt}=\sum_iV_i\frac{\partial\rho_i}{\partial t}=-\sum_i{V_i\nabla_i\cdot\rho\mathbf{u}}=-\sum_i{\sum_{i^\prime}\mathbf{S}_{ii^\prime}\cdot\left(\rho\mathbf{u}\right)_{ii^\prime}}=0,$$ 
since two contributions from each $ii^\prime$-th face flux, $\left(\mathbf{S}_{ii^\prime}+\mathbf{S}_{i^\prime i}\right)\cdot\left(\rho\mathbf{u}\right)_{ii^\prime}$, vanish because ${\bf S}_{ii^\prime}=-{\bf S}_{i^\prime i}$. Analogously, the momentum and energy conservation is achieved. The Lagrangian step describes the change in fluid state in the control volume due to the motion re-distributing the fluid between different volumes. This effect is directly comparable with the particle coordinate advance in the PIC algorithms:
\begin{equation}\label{eq:PICCoordinates}
\mathbf{x}^{N+\frac12}_p=\mathbf{x}^{N-\frac12}_p+\Delta t\mathbf{u}^N_p,
\end{equation}
where $\mathbf{x}_p$ and $\mathbf{u}_p$ are the coordinate and velocity vectors for particle, enumerated with the index, $p,$ and the time indexes (semi-integer for coordinate, integer for velocity) emphasize the leapfrog time-centering manner of PIC algorithms. The {\it ballistic} motion of particles with unchanged velocity in Eq.~(\ref{eq:PICCoordinates}) corresponds to the CFD transport in Eq.~(\ref{eq:LagrangianStep}): the fluid velocity varies not because any particle velocity changes in time, but because faster or slower particles are transported from the neighboring domain. The mass and momentum conservation in the CFD scheme complies with the conservation of number of particles and momentum of each particle in the course of ballistic propagation in Eq.~(\ref{eq:PICCoordinates}).

\textbf{Step 2:} At the Eulerian stage, following \cite{Belotserkovsky:71} the content of the  control volume is treated as a "coarse particle", located at the cell center, $\mathbf{x}_p=\mathbf{x}_i$. The force, exerted by a particle of volume $V_p$ in the pressure field, $P(\mathbf{x})$, equals $\mathbf{F}_p(\mathbf{x}_p)=-V_p\left[\nabla P(\mathbf{x})\right]_{\mathbf{x}=\mathbf{x}_p}$, therefore, the particle acceleration in terms of its mass, $m_p$ in the leapfrog manner PIC scheme, is calculated as follows. 
\begin{equation}
\label{eq:PICVelocities}
\mathbf{u}_p^{N+1}=\mathbf{u}_p^{N}+\frac{\Delta t}{m_p}\mathbf{F}_p\left(\mathbf{x}^{N+1/2}_p\right).
\end{equation}
While in the PIC method the effect of force field is accounted for each of numerous particles and depends on the particle location, in the FLIC scheme Eq.~(\ref{eq:PICVelocities}) is applied only to cell-centered coarse particle of mass, $m_p=\rho_iV_i$, thus giving $\rho_i\frac{\partial \mathbf{u}_i}{\partial t}=-\nabla_iP=-\sum_{i^\prime}\mathbf{S}_{ii^\prime}P_{ii^\prime}$. Analogously, the total work done by pressure changes the energy, so that one can formulate the Eulerian stage of the coarse particle method as follows:
\begin{equation}
\label{eq:EulerianStep}
\rho_i\frac{\partial}{\partial t}\left(
\begin{array}{c}\mathbf{u}\\ \frac{u^2}2+\frac{P}{\rho_i\left(\gamma-1\right)}\end{array}\right)+\nabla_i\cdot P\left(
\begin{array}{c}\cal{I}\\\mathbf{u}\end{array}\right)=0,
\end{equation}
with $\cal{I}$ being the unit tensor and density $\rho_i$  not evolving within this stage. The  Lagrangian and Eulerian steps of the FLIC (coarse particle) method both respect the hydrodynamic conservation laws. 
\subsection{{\color{black}Coarse} Particle Magnetohydrodynamics}\label{sec:mhd}
\begin{figure*}[htbp]
\begin{center}
\includegraphics[width=\textwidth]{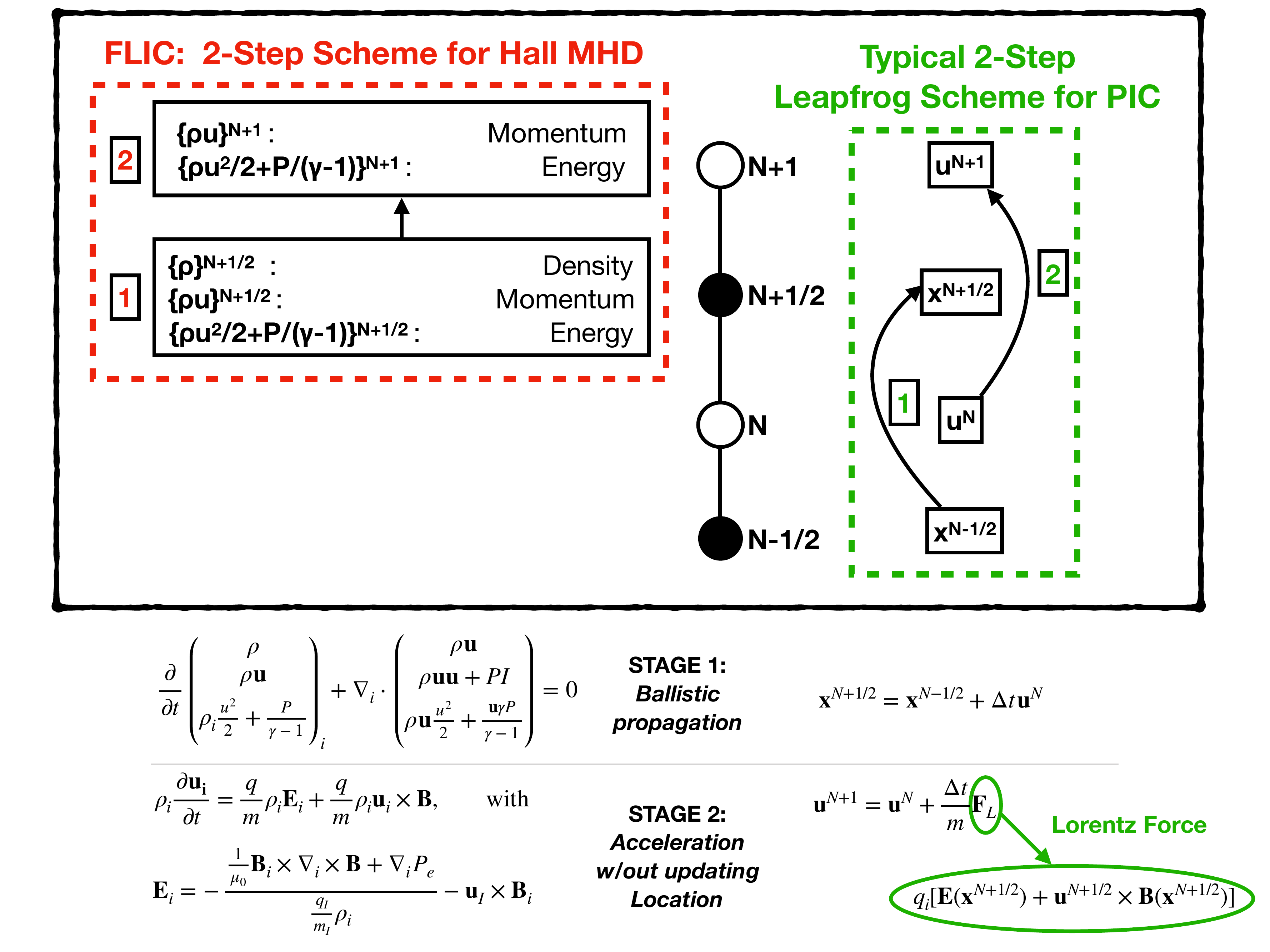}  \caption{Schematic summarizing the similarities in the numerical methods of hybrid and MHD models. When electromagnetic fields are important, the particle acceleration is due to the Lorentz force and not the pressure gradient. Both use the generalized Ohm's law and identical field.}
  \label{fig:mhd}
  \end{center}
  \end{figure*}

Following the same rationale, as the one presented in the previous section for the purely hydrodynamic case, here we generalize the justification to combine hybrid and MHD within a single code when electromagnetic fields are important. These two physical models are similar in that they both use the generalized Ohm's law and their respective field equations are identical. Both two-step schemes are shown in Fig.~\ref{fig:mhd}

The MHD model describes the motion of highly conducting fluids or plasmas in the magnetic field. Such motion within the hybrid model is treated as the motion of electrically charged ion (macro)particles in self-consistent electromagnetic fields. Particularly, a thermal motion of ions is simulated via a spread in the particle velocities, such that the particle velocities, $\mathbf{u}_
p$, differ from the averaged local velocity within a control volume, $\mathbf{u}_i$. Therefore, in generalizing the FLIC scheme for MHD it is natural to account for the thermal motion of particles by including the pressure effect (which is nothing but the momentum transfer in the course of a thermal motion) to the Lagrangian step:
\begin{equation}\label{eq:LagrangianMHD}
\frac{\partial}{\partial t}\left(
\begin{array}{c}\rho\\\rho\mathbf{u}\\ \rho\frac{u^2}2+\frac{P}{\left(\gamma-1\right)}\end{array}\right)_i+\nabla_i\cdot\left(
\begin{array}{c}\rho\mathbf{u}\\\rho\mathbf{u}\mathbf{u}+P\cal{I}\\ \rho\mathbf{u}\frac{u^2}2+\frac{\mathbf{u}\gamma P}{\left(\gamma-1\right)}\end{array}\right)=0.
\end{equation}
Eq.~(\ref{eq:LagrangianMHD}) integrates the effect of individual particle motion. In the coupled hybrid-MHD model, concurrently with the stage \ref{eq:LagrangianMHD} in the global MHD domain, the particles of the hybrid scheme move according to Eq.~(\ref{eq:PICCoordinates}).

%
For the Eulerian stage, here, we discuss only the momentum equation, in which the flux tensor, $\Pi$, may be found as the difference between the total momentum flux and that one accounted for in Eq.~(\ref{eq:LagrangianMHD}):
\begin{equation}\label{eq:MomentumMHD}
\rho_i\frac{\partial\mathbf{u}_i}{\partial t}+\nabla_i\cdot\Pi=0,\qquad\Pi=\left(P_e+\frac{B^2}{2\mu_0}\right)\cal{I}-\frac{\mathbf{B}\mathbf{B}}{\mu_0}, 
\end{equation}
with the conservative formulation for the flux divergence, $\nabla_i\cdot\Pi=\sum_{i^\prime}\mathbf{S}_{ii^\prime}\cdot\Pi_{ii^\prime}/V_i$. Eq.~(\ref{eq:MomentumMHD}) accounts for an electromagnetic force effect on ion fluid, which for an individual ion particle is described by
Eq.~(\ref{eq:PICVelocities}):
\begin{equation}
\label{eq:PICVelocitiesEM}
\mathbf{u}_p^{N+1}=\mathbf{u}_p^{N}+\Delta t\frac{q_p}{m_p}
\left[\mathbf{E}^{N+\frac12}\left(\mathbf{x}_p^{N+\frac12}\right)+\mathbf{u}^{N+\frac12}_p\times\mathbf{B}^{N+\frac12}\left(\mathbf{x}_p^{N+\frac12}\right)\right],
\end{equation}
where $q_i$ is the ion charge, $\mathbf{E}$, $\mathbf{B}$ are electric and magnetic fields interpolated to the ion location, $\mathbf{x}^{N+\frac12}_p$. Eq.~(\ref{eq:MomentumMHD}) may be interpreted as Eq.~(\ref{eq:PICVelocitiesEM}) applied to the cell-centered coarse particle of the mass, $m_i=\rho_i V_i$, and charge, $q_i=\frac{q_p}{m_p}\rho_i V_i$, such that $\frac{\partial\mathbf{u}_i}{\partial t}=\frac{q_i}{m_i}\left(\mathbf{E}_i+\mathbf{u}_i\times\mathbf{B}_i\right)$, with the discrete {\it generalized Ohm's law},
\begin{equation}\label{eq:OhmsLawDiscrete}
\mathbf{E}_i=
-\frac{\sum_{i^\prime}{\Pi_{ii^\prime}\cdot \mathbf{S}_{ii^\prime}}}{\tilde{\rho}_{i}V_i}
-\mathbf{u}_i\times\mathbf{B}_i,
\end{equation}
expressing the cell-centered electric field, $\mathbf{E}_i$, in terms of the cell-centered magnetic field, $\mathbf{B}_i$, the momentum flux divergence combining $\mathbf{J}\times\mathbf{B}$ force and an electron pressure gradient, the ion charge density, $\tilde{\rho}_{i}=\frac{q_i}{m_i}\rho_{i}$ expressed in terms of the cell-centered mass density, $\rho_i$, as well as the ion velocity, $\mathbf{u}_i$. The differential form of Eq.~(\ref{eq:OhmsLawDiscrete}) may be derived using Eq.~(\ref{eq:MomentumMHD}) for momentum flux and equation, $\nabla\cdot\mathbf{B}=0$:
\begin{equation}\label{eq:GenOhmLaw}
\mathbf{E}=-\frac{\frac1{\mu_0}\mathbf{B}\times\nabla\times\mathbf{B}+\nabla P_e}{\tilde{\rho}}-\mathbf{u}\times\mathbf{B}.  
\end{equation}
Alternatively, Eq.~(\ref{eq:GenOhmLaw}) may be derived from the momentum equation for \textit{massless} electrons ($\rho_e=0$), $
        \rho_e \frac{d\mathbf{u_e}}{dt}=0=\tilde{\rho}_e\mathbf{E}+\mathbf{J_e}\times\mathbf{B}-\nabla P_e$,
assuming quasi-neutrality, $\tilde{\rho}+\tilde{\rho}_e=0$, and using the Ampere's law with no displacement current, $
\mathbf{J}_e+\mathbf{J}=\frac{1}{\mu_0}\nabla\times\mathbf{B}$, to relate the electron current, $\mathbf{J}_e$ to the ion current, $\mathbf{J}=\tilde{\rho}\mathbf{u}$. The total momentum conservation in the course of Eulerian step, is easy to demonstrate:
$$
\frac{d}{dt}\sum_i\rho_i\mathbf{u}_iV_i=\sum_i\rho_i\frac{\partial\mathbf{u}_i}{\partial t}V_i=\sum_i\tilde{\rho}V_i\left(-\frac{\sum_{i^\prime}{\Pi_{ii^\prime}\cdot \mathbf{S}_{ii^\prime}}}{\tilde{\rho}_{i}V_i}
-\mathbf{u}_i\times\mathbf{B}_i+\mathbf{u}_i\times\mathbf{B}_i\right)=-\sum_i\sum_{i^\prime}{\Pi_{ii^\prime}\cdot \mathbf{S}_{ii^\prime}}=0.
$$
So far, we oversimplify the coarse particle method for MHD by implicit assumption that only a single sort of ions is present in a plasma, so that the charge-to-mass ratio is the same for all ion particles. In this case the terms, $\pm\mathbf{u}_{i}\times\mathbf{B}_i$, in the ion acceleration and electric field cancel each other, so that the coarse ion velocity does not affect its acceleration. More typical for space plasmas is the presence of several sorts of ions, with different $q_s/m_s$ ratios, different velocities, $\mathbf{u}_s$ and densities, $\rho_s$, where index, $s$, enumerates ion sorts \cite[see][for multi-ion MHD simulation results]{Huang2016,Huang2018}. For this case the coarse particle for each sort of ions is needed, thus resulting in the equations for Eulerian stage as follows:
\begin{equation}\label{eq:multiion}
\rho_s\frac{\partial\mathbf{u}_{si}}{\partial t}=\frac{q_s}{m_s}\rho_s\left(\mathbf{E}_i+\mathbf{u}_{si}\times\mathbf{B}_i\right),
\end{equation}
\begin{equation}\label{eq:OhmsLawDiscrCons}
\mathbf{E}_i=
-\frac{\sum_{i^\prime}{\Pi_{ii^\prime}\cdot \mathbf{S}_{ii^\prime}}}{\tilde{\rho}_{i}V_i}
-\frac{\mathbf{J}_{i}\times\mathbf{B}_i}{\tilde{\rho}_i},
\end{equation}
the {\it auxiliary} quantities, the ion charge and current densities, being expressed  via the MHD variables:
\begin{equation}\label{eq:rhoJ}
\tilde{\rho}_{si}=\sum_s\frac{q_s}{m_s}\rho_{si}, \qquad \mathbf{J}_i=\sum_s\frac{q_s}{m_s}\rho_{si}\mathbf{u}_{si}.    
\end{equation} 
Within the multi-coarse-particles conservative MHD scheme \ref{eq:multiion}-\ref{eq:OhmsLawDiscrCons}, the momentum flux divergence (or, force), 
$\sum_{i^\prime}\Pi_{ii^\prime}\cdot \mathbf{S}_{ii^\prime}/V_i$, is distributed over different sorts of ions, the fraction per the given sort, $s$, being equal to $\frac{q_s}{m_s}\rho_{si}/\tilde{\rho}_i$. The sum of these fractions is equal to one, $\sum_s\frac{q_s}{m_s}\rho_{si}/\tilde{\rho}_i=\tilde{\rho}_i/\tilde{\rho}_i=1$, allowing for the total momentum conservation. The velocity of the given sort of ions both is directly included into the Lorentz force and contributes to the electric field via the $-\mathbf{J}\times\mathbf{B}$ term. The sum of velocity dependent forces over the ion sorts, however, equals zero and does not break the total momentum conservation.

 The magnetic field is advanced in time through the Faraday law:
    \begin{equation}\label{eq:faraday}
        \frac{\partial \mathbf{B}}{\partial t}= -\nabla\times\mathbf{E}.
    \end{equation}
    If the full electric field from Eq.~(\ref{eq:GenOhmLaw}) is used in Eq.~(\ref{eq:faraday}), the Hall effect is thus accounted. In the ideal MHD approximation the electric field in Eq.~(\ref{eq:faraday}) is taken to be $-\mathbf{J}\times \mathbf{B}/\tilde{\rho}$.  The electron pressure $P_e$,  affects the ions through electric field. It is assumed to be isotropic and can be solved from the equation for the internal energy of electrons, with the adiabatic index, $\gamma=5/3$:
        \begin{equation}\label{eq:efluidenergy}
        \frac{\partial P_e}{\partial t}+\mathbf{u_e}\cdot\nabla P_e + \gamma P_e\nabla\cdot\mathbf{u_e}=0,\qquad \mathbf{u}_e= \frac{\mathbf{J}}{\tilde{\rho}} - \frac{\nabla\times \mathbf{B}}{\mu_0 \tilde{\rho}}.
    \end{equation}
Again, in the ideal MHD it is natural to approximate, $\mathbf{u}_e= \frac{\mathbf{J}}{\tilde{\rho}}$. Eqs.~(\ref{eq:faraday}), (\ref{eq:efluidenergy}) may be solved both for FLIC-MHD and hybrid models, in the latter case the cell-centered moments, $\rho_{si}$, $\rho_{si}\mathbf{u}_{si}$, of the distribution functions for different sorts of ions need to be collected, to represent the corresponding MHD variables.
\section{Hybrid Code}\label{sec:hybrid}
 The most common ingredients in a hybrid code are the following \cite[see, e. g.][]{Holmstrom:09}.
 \begin{enumerate}
     \item Fields are interpolated at the particle positions, ${\bf x}^{N+1/2}$, to calculate the acceleration, $\mathbf{a}^{N+{1/2}}$, for all ions, and advance their velocities: $\mathbf{u}^{N+1} = \mathbf{u}^{N} + \Delta t \mathbf{a}^{N+{1/2}}$ .
     \item The moments of distribution functions are calculated back on the grid to calculate the charge and current densities and electric field. To do this, one needs to have particle coordinates and velocities at the same time level/sub-level.
     \item However, particle positions $\mathbf{x}^{N+1/2}$ and velocities $\mathbf{u}^N$ are leapfrogged in time. To evaluate the current density at level $N$, one can just use the linear time interpolation, ${\bf x}^N=\frac12\left({\bf x}^{N-1/2}+{\bf x}^{N+1/2}\right)$, but for ${\bf u}^{N+1/2}$ such estimate is not as straightforward, since the $\mathbf{u}^{N+1}_i$ is not yet known.
     \item To calculate accelerations, the fields need to be calculated at the half time-step. The magnetic field update, $\mathbf{B}^N\rightarrow \mathbf{B}^{N+\frac12}$, may be done explicitly  through the \textit{Faraday law} Eq.~(\ref{eq:faraday}). The electric field $\mathbf{E}^{N+\frac12}\leftarrow\mathbf{B}^{N+\frac12}_i,P_e^{N+\frac12},\tilde{\rho}^{N+\frac12}_i,\mathbf{J}^{N+1/2}_i$, is solved from the Ohm's generalized law, Eq.~(\ref{eq:GenOhmLaw}).
To do this, one needs to know the charge and current density, which is the most difficult to get, at half time-step.
    \item Faraday's law is also used to  advance the magnetic field $\mathbf{B}$ to the full time-step $\mathbf{B}^{N}\rightarrow \mathbf{B}^{N+1}$, as long as the electric field,  at half time-step is known. 
 \end{enumerate}

What will make a hybrid code more useful will be its ability to self-consistently calculate the electric field $\mathbf{E}^{{N+1/2}}$, and electric current density, $\mathbf{J}^{N+1/2}$, in the most efficient way.
One of the possible solutions to the problem is to extrapolate not the particle velocities, but their integral characteristic, ion current density, since this is all we need to determine the electric field. Here, we use the Current Advance Method (CAM) by \cite{Matthews:94}, which explicitly computes the ion current $\mathbf{J}^{{N+\frac12}}$ by using a mixed level evaluation of the electric field $\mathbf{E}^{\star}$ and the ion current $\mathbf{J}^{\star}$. With the superscript ($^\star$) we denote the "mixed level" evaluations, which are calculated in terms of $x^{N+1/2}_p,\mathbf{u}^{N}_p$. The CAM is identical to the Eulerian step of FLIC MHD as we explain here. 

\subsection{Time Advance Algorithm}\label{sec:CAM}
\begin{figure*}[htbp]
  \begin{center}
  \includegraphics[width=\textwidth,trim={2cm 8.5cm 2cm 1cm},clip]{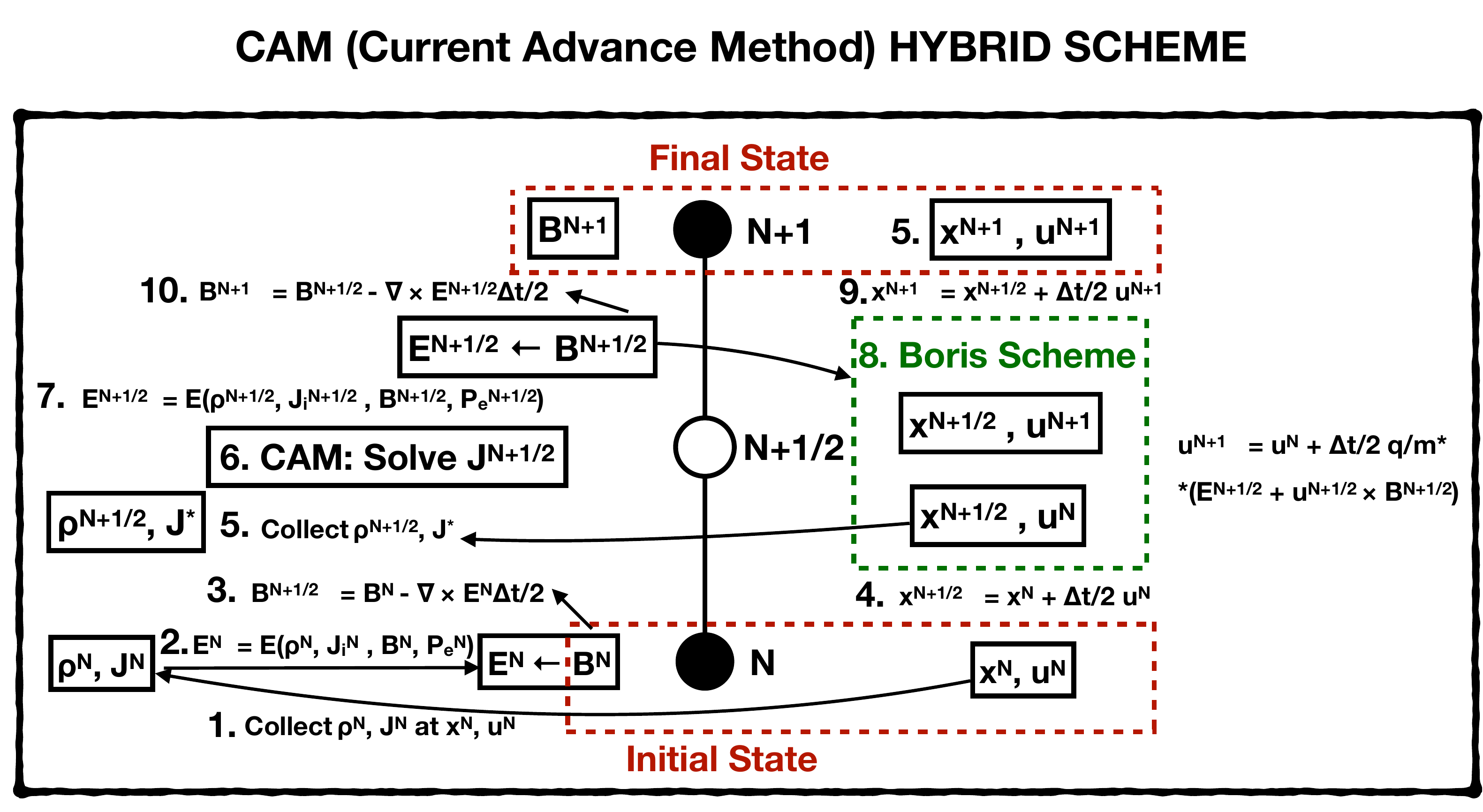}  \caption{Schematic of the time advancement method of the hybrid code to advance the electric field $\mathbf{E}^{N}\rightarrow \mathbf{E}^{{N+1}}$ as adapted from \cite{Matthews:94} in SWMF.}
   \label{fig:diagram}
   \end{center}
   \end{figure*}

The hybrid algorithm is explained in Fig. \ref{fig:diagram}. At stage 1, we collect moments of the distribution functions, 
$$
\rho^N_{si}=
\frac1{V_i}\sum_{p_s}{m_{p_s}w^N_{p_si}\left(\mathbf{x}^{N}_{p_s},\mathbf{x}_i\right)},
\qquad
\left(\rho\mathbf{u}\right)_{si}^N=\frac1{V_i}
\sum_{p_s}{m_{p_s}w^N_{p_si}\left(\mathbf{x}^{N}_{p_s},\mathbf{x}_i\right)\mathbf{u}^N_{p_s}},
$$ 
for each cell, enumerated by index $i$, and for each sort of ions, enumerated by index, $s$.
Herewith, $V_i$ is the volume of the $i$-th cell, $m_{p_s}$ is the mass of macroparticle of $s$-th sort , enumerated with the index, $p_s$, and $w^{N}_{p_si}\left(\mathbf{x}^{N}_{p_s},\mathbf{x}_i\right)$ is the fraction of the mass of $p_s$-th particle related to the $i$th cell, which depends on the relative location of the particle, $\mathbf{x}^{N}_{p_s}$, with respect to the cell centers, $\mathbf{x}_i$. The sum of these coefficients for a given particle equals one: $\sum_iw_{p_si}=1$, which means that the total of the fractions of the mass decomposed over all the cells equals the total mass of the macro-particles. Such weight coefficients are also used to interpolate the field acting on a given particle according to Eqs.~(\ref{eq:EBp}). The algorithm to calculate these weight coefficients for {\it adaptive} grids may be found in \citep{Borovikov2015}.

Then, we calculate ion charge and current densities using Eq.~(\ref{eq:rhoJ}), which allows us to get the electric field (stage 2), 
$\mathbf{E}_i^N\leftarrow\mathbf{B}^{N}_i,P_e^{N},\tilde{\rho}^{N}_i,\mathbf{J}^N_i$, and (stage 3) to
advance the magnetic field through a half time step:
\begin{equation}\label{updateB}
\mathbf{B}^{N+\frac12}_{i}=\mathbf{B}^{N}_{i}-
\frac{\Delta t}2\left(\nabla\times\mathbf{E}^{N}\right)_{i},
\end{equation}
and solve $P_e^{N+1/2}$ from Eq.~(\ref{eq:efluidenergy}). Then we pass over particles, and update their coordinates (stage 4):
\begin{equation}\label{eq:x}
\mathbf{x}_p^{N+\frac12}=\mathbf{x}_p^{N}+\frac12\Delta t\mathbf{u}^{N}_p.
\end{equation}
{\color{black}Being equivalent to a "theoretical" leapfrog formulation,  $\mathbf{x}_p^{N+1/2}=\mathbf{x}_p^{N-1/2}+\Delta t\mathbf{u}^{N}_p$, the  use of  intermediate particle location, $\mathbf{x}_p^{N}=\frac12\left(\mathbf{x}_p^{N+\frac12}+\mathbf{x}_i^{N-\frac12}\right)$, is advantageous, since it does not require time steps to be equal}. Now, following \cite{Matthews:94}, we collect the moments of the ion distribution functions, which combine the {\it updated} particle coordinates and {\it outdated} velocities (stage 5):
\begin{equation}\label{eq:rhoustar}
\rho^{N+1/2}_{si}=
\frac1{V_i}\sum_{p_s}{m_{p_s}w^{N+1/2}_{p_si}\left(\mathbf{x}^{N+1/2}_{p_s},\mathbf{x}_i\right)},
\qquad
\left(\rho\mathbf{u}\right)_{si}^*=\frac1{V_i}
\sum_{p_s}{m_{p_s}w^{N+1/2}_{p_si}\left(\mathbf{x}^{N+1/2}_{p_s},\mathbf{x}_i\right)\mathbf{u}^N_{p_s}},
\end{equation}
which allows us to calculate the charge density $\tilde{\rho}^{N+1/2}$ and predict the current density $\mathbf{J}^*$ using Eq.~(\ref{eq:rhoJ}). This stage of hybrid algorithm is analogous to the Lagrangian step of the FLIC MHD scheme. 
   
The heart of the CAM approach (stage 6) is in the use of the electric field calculated with thus approximated current density, $
\mathbf{E}_i^* \leftarrow \mathbf{B}^{N+\frac12}_i,P_e^{N+\frac12},\tilde{\rho}^{N+\frac12}_i,\mathbf{J}^*_i$.
which then allows us to extrapolate the electric current at the half time-step. The explicit 
formulae by \cite{Matthews:94} for direct extrapolation of current can be conveniently rewritten in terms of extrapolation of the distribution function moments:  
\begin{equation}\label{eq:CAM}
\left(\rho\mathbf{u}\right)_{si}^{N+1/2}=\left(\rho\mathbf{u}\right)_{si}^*+\frac{\Delta t}{2}\frac{q_s}{m_s}\left[\rho^{N+1/2}_{si} \mathbf{E}^{\star}_i + \left(\rho\mathbf{u}\right)_{si}^* \times \mathbf{B}^{N+1/2}_i\right],
\end{equation}
which is identical to advancing Eqs.~(\ref{eq:multiion}) in Eulerian stage of the FLIC MHD scheme through a half time-step using \textit{explicit} approximation for particle velocity in the RHS (an implicit scheme to solve $J^{N+1/2}_i$ may be found in Appendix). 
The current density, $\mathbf{J}^{N+1/2}_i$ is calculated from Eqs.~(\ref{eq:rhoJ}), (\ref{eq:CAM}). By solving the electric field at half time-step (this is stage 7 of the algorithm):
\begin{equation}\label{eq:ECAM}
\mathbf{E}_i^{N+1/2} \leftarrow \mathbf{B}^{N+\frac12}_i,P^{N+1/2}_e,\tilde{\rho}^{N+\frac12}_i,\mathbf{J}^{N+1/2}_i,
\end{equation}
we are now prepared to push particles (stages 8-9). The fields at the particle position 
are interpolated: 
\begin{eqnarray}\label{eq:EBp}
\{{\bf E},{\bf B}\}\left({\bf x}^{N+1/2}_p\right)=\sum_iw^{N+1/2}_{pi}\left(\mathbf{x}_p^{N+1/2},\mathbf{x}_i\right)\{{\bf E},{\bf B}\}^{N+1/2}_i
\end{eqnarray}
Eq.~(\ref{eq:PICVelocitiesEM}) for the particle momentum is then solved using the Boris scheme \citep{Birdsall05}:
\begin{eqnarray}\label{eq:Boris}
\hat{\bf u}^{N+\frac12}_p&=&{\bf u}^{N}_p+\frac{q_p\Delta t}{2m_p}{\bf E}({\bf x}^{N+\frac12}_p),\qquad\quad
\mathbf{u}^{N+\frac12}_p=\frac{\hat{\bf u}^{N+\frac12}_p+
\hat{\bf u}^{N+\frac12}_p\times\frac{q_p\Delta t}{2m_p}
{\bf B}({\bf x}^{N+\frac12}_p)}
{1+\left[\frac{q_p\Delta t}{2m_p}
{\bf B}({\bf x}^{N+\frac12}_p)\right]^2},\\
{\bf u}^{N+1}_p&=&\mathbf{u}^{N}_p+\frac{q_p\Delta t}{m_p}{\bf E}({\bf x}^{N+\frac12}_p)+{\bf u}^{N+\frac12}_p\times\frac{q_p\Delta t}{m_p}
{\bf B}({\bf x}^{N+\frac12}_p),\nonumber
\end{eqnarray}
where the denominator in Eq.~\ref{eq:Boris} is chosen in such a manner that the magnetic 
field does not affect the particle energy. 
With the updated velocity, the particle location is advanced through a half time-step:
${\bf x}^{N+1}_p={\bf x}^{N+1/2}_p+\frac{\Delta t}2{\bf u}^{N+1/2}_p$. In the FLIC MHD scheme
this stage corresponds to advancing Eq.~(\ref{eq:multiion}) through the whole time step followed by the Lagrangian advance through the half time-step.

At the last stage, the magnetic field, $\mathbf{B}^{N+1}$, and electron pressure, $P_e^{N+1/2}$, should be advanced through half time-step using Eqs.~(\ref{eq:faraday}-\ref{eq:efluidenergy}). We finish the update for this time step and proceed to the next one. 

\subsection{Momentum Conservation}
Now, we can analyze a change in the total particle momentum, $\sum_pm_p\mathbf{u}^N_p$, resulting from the field effect, by multiplying Eqs.~(\ref{eq:PICVelocities}) by the particle mass and then summing them up over particles. In turn, in the expressions for fields, Eqs.~(\ref{eq:EBp}), the sums over $i$ are present. On changing an order of summation, the increment in the total momentum can be represented as the sum over cells:
\begin{equation}\label{eq:momentum}
\sum_pm_p\mathbf{u}^{N+1}_p-\sum_pm_p\mathbf{u}^{N}_p
=\Delta t\sum_i{V_i\left(\tilde{\rho}^{N+\frac12}_{i}\mathbf{E}^{N+\frac12}_{i}+\mathbf{J}^{N+\frac12}_i\times\mathbf{B}^{N+\frac12}_i\right)}.
\end{equation}
 With an arbitrary discretization of the electric field in Eq.~(\ref{eq:GenOhmLaw}), the RHS of Eq.~(\ref{eq:momentum}) does not necessarily vanish, so that the total particle momentum is not conserved. However,  Eq.~(\ref{eq:momentum})  gives us a hint that the momentum conservation may be achieved, if the force acting on the particles obeys the formulation as follows:
$$
V_i\left(\rho^{N+\frac12}_{i}\mathbf{E}^{N+\frac12}_{i}+\mathbf{J}^{N+\frac12}_i\times\mathbf{B}^{N+\frac12}_i\right)=-\sum\limits_{i^\prime}{\left(\Pi_{ii^\prime}\cdot \mathbf{S}_{ii^\prime}\right)^{N+1/2}},
$$
or, equivalently, if the discrete generalized Ohm's law given by Eq.~(\ref{eq:OhmsLawDiscrCons}) is chosen for electric field, the same is that in the FLIC MHD.
With this choice, Eq.~(\ref{eq:momentum}) demonstrates the total momentum conservation:
$$
 \sum_pm_p\mathbf{u}^{N+1}_p-\sum_pm_p\mathbf{u}^{N}_p
=-\Delta t\sum_i{
\sum\limits_{i^\prime}
\left(\Pi_{ii^\prime}\cdot \mathbf{S}_{ii^\prime}
\right)^{N+1/2} }=0,
 $$
the double sum vanishes by the reason explained above. 

Thus, both for FLIC MHD and for hybrid scheme the cell-centered electric field acting on particles should be expressed in terms of the momentum flux divergence, further contributing to a deep similarity of these computational models. We notice above that the CAM algorithm of hybrid scheme is equivalent to the Eulerian step Eq.~(\ref{eq:multiion}) of the FLIC MHD. Common numerical schemes may be also used,  to update the electron pressure and magnetic field, whose evolution in both models is governed by the distribution function moments, $\rho_{si}$ and $(\rho\mathbf{u})_{si}$. With these regards, the numerical framework, which solves numerically Eqs.~(\ref{eq:multiion}), (\ref{eq:rhoJ}), (\ref{eq:faraday}), (\ref{eq:efluidenergy}), with proper time staging and calculates $\sum_{i^\prime}
\left(\Pi_{ii^\prime}\cdot \mathbf{S}_{ii^\prime}
\right)^{N+1/2}$ and then the half time-step electric field $\mathbf{E}^{N+1/2}$, provides a perfect environment to integrate numerically the MHD (or Hall MHD) or hybrid models, or to combine both models in different parts of the computational domain, or solve some sorts of ions as fluids and other ones as particles. By all means, the numerical solution conserves mass and momentum. The energy conservation will be addressed elsewhere.   
\section{Conclusions}\label{sec:discussion}
The solar corona is a highly dynamic region, with physical processes that involve both macroscopic and particle scales. The underlying physics require a combination of both MHD and particle models to be modeled adequately. To tackle this problem, we implemented a new coupled scheme within the SWMF, which allows to switch blocks in the block-adaptive grids from fluid MHD to hybrid simulations, without modifying the self-consistent computation of the electromagnetic fields acting on fluids (in MHD simulations) or charged ion macroparticles (in hybrid simulations). In this way, the hybrid scheme can provide a refined description of specified places of interest without compromising the efficiency of the global MHD code.
The implemented hybrid algorithm conserves the total momentum and achieves high accuracy discretization for Ohm's law. The coupled MHD--hybrid scheme can be applied for multiple fluids and may account for different kinds of ions, including test particles (neglect their current) and/or full hybrid particles.
Currently, we are bench-marking the code by performing a suite of tests for different plasma conditions including multi-fluid setups, that we will present in a more detailed future publication.

\section*{Acknowledgements}
SPM was supported by NASA Living with a Star grant number NNX16AC11G. IVS was supported by INSPIRE NSF grant PHY-1513379 We would also like to acknowledge high-performance computing support from: (1) Yellowstone (ark:/85065/d7wd3xhc) provided by NCAR's Computational and Information Systems Laboratory, sponsored by the National Science Foundation, and (2) Pleiades operated by NASA's Advanced Supercomputing Division.  JJD
was funded by NASA contract NAS8-03060 to the {\it Chandra X-ray Center}.
\appendix
\section{Numerical Schemes for Eqs.~(\ref{eq:multiion}), (\ref{eq:OhmsLawDiscrCons}) for CAM or Eulerian Step of FLIC MHD}
Here, we describe a numerical scheme to solve Eqs.~(\ref{eq:multiion}), (\ref{eq:OhmsLawDiscrCons}), which in the original CAM algorithm by \cite{Matthews:94} are treated by explicit scheme  (\ref{eq:CAM}). However, in the works on multi-ion MHD by \cite{Huang2016,Huang2018} the $\mathbf{J}\times\mathbf{B}$ force for each ion sort with stiff dependence on velocities and magnetic field have been handled using implicit scheme, in these equations. The semi-implicit  Crank-Nicolson  scheme proposed here is unconditionally stable, more accurate than the implicit scheme, and can be explicitly resolved. 

Eqs.~(\ref{eq:multiion}), (\ref{eq:OhmsLawDiscrCons}) relate quantities for $i$-th  control volume, therefore, we omit index $i$ for simplicity, use full time derivatives instead of the partial ones and re-written the equations as follows:
\begin{eqnarray}\label{eq:appgov}
    \frac{d\left(\rho\mathbf{u}\right)_s}{dt}=\kappa_s\mathbf{f}+\Omega_s\left(\rho\mathbf{u}\right)_s\times\mathbf{b}&-&\kappa_s\sum_{s^\prime}\Omega_{s^\prime}\left(\rho\mathbf{u}\right)_{s^\prime}\times\mathbf{b},\\
 \mathbf{f}=-\frac{\sum_{i^\prime}\left(\Pi_{i^\prime}\cdot\mathbf{S}_{i^\prime}\right)^{N+1/2}}{V},&\quad&\Omega_s=\frac{q_s}{m_s}B,\qquad \kappa_s=\frac{q_s}{m_s}\frac{\rho^{N+1/2}_s}{\tilde{\rho}^{N+1/2}},\nonumber
\end{eqnarray}
where $B=|\mathbf{B}^{N+1/2}|$ and $\mathbf{b}=\mathbf{B}^{N+1/2}/B$ are the magnitude of the magnetic field and its direction vector, $\kappa_s$ is the relative charge density for $s$-th ion, $\Omega_s$ is a gyrofrequency for $s$-th ion, and $\mathbf{f}$ is the force due to the momentum flux divergence. Initial ion velocities are known: $\left(\rho\mathbf{u}\right)_s|_{t=t^N}=\left(\rho\mathbf{u}\right)^N_s$. 

First, we solve Eqs.~(\ref{eq:appgov}) for the motion along the magnetic field: $\left(\rho\mathbf{u}\right)^{N+1}_s\leftarrow\mathbf{b}
\left\{\mathbf{b}\cdot\left[\left(\rho\mathbf{u}\right)^{N}_s+
\Delta t\,\kappa_s\mathbf{f}\right]\right\}$,
and then keep only perpendicular vector components in the initial condition:
$\left(\rho\mathbf{u}\right)^{N}_s\leftarrow\left(\rho\mathbf{u}\right)^{N}_s-\mathbf{b}\left[\mathbf{b}\cdot\left(\rho\mathbf{u}\right)^{N}_s\right]
$, and in the force: $\mathbf{f}\leftarrow\mathbf{f}-\mathbf{b}\left(\mathbf{b}\cdot\mathbf{f}\right)$. 

In the Cartesian coordinate frame with $z$-axis aligned with $\mathbf{b}$ the vector of perpendicular velocity for each sort of ions can be equivalently represented as \textit{complex} variable: $Z_s=\left(\rho u_x\right)_s+j\left(\rho u_y\right)_s\leftrightarrow\left(\rho\mathbf{u}\right)_s$, $j$ being the imaginary unit. Analogously, we can introduce the complex representative for perpendicular force, $F=f_x+jf_y\leftrightarrow\mathbf{f}$. It is easy to check that $-j Z_s\leftrightarrow\left(\rho\mathbf{u}\right)_s\times\mathbf{b}$. 
On combining $S$ complex velocities into the $S$-component column vector $\mathbf{Z}=\left\{Z_s\right\}$, and on introducing the $S$-component column vector, $\mathbf{K}=\left\{\kappa_s\right\}$, and $S$-component line vector, $\mathbf{O}^T=\left\{\Omega_s\right\}^T$, as well as the diagonal $S$-by-$S$ matrix, $\mathrm{diag}\{\Omega_s\}$, one can derive Eqs.~(\ref{eq:appgov}) for the perpendicular motion in a vector-matrix form:
\begin{equation}\label{eq:appvect}
   \frac{d\mathbf{Z}}{dt}=F\mathbf{K}-jA\cdot\mathbf{Z},\qquad A=\mathrm{diag}\{\Omega_s\}-\mathbf{K}\otimes\mathbf{O}^T,
\end{equation}
$S$ being the number of sorts of ions with different charge-to-mass ratios. The $S$-by-$S$ matrix, $A$, defined in Eq.~(\ref{eq:appvect}), can be \textit{decomposed},
by constructing a set of eigenvalues, $\omega$, right (column) eigenvectors, $\mathbf{R}=\left\{r_s\right\}$, and left (line) eigenvectors, $\mathbf{L}^T=\left\{l_s\right\}^T$, such that for right eigenvectors:
$\omega\mathbf{R}=A\cdot\mathbf{R}$, and for left eigenvectors: $\omega\mathbf{L}^T=\mathbf{L}^T\cdot A$.
The eigenvectors,  
\begin{equation}\label{eq:appeigvec}
\mathbf{R}=\left\{\frac{\kappa_s}{\Omega_s-\omega}\right\}, \qquad \mathbf{L}^T=\left\{\frac{\Omega_s}{\Omega_s-\omega}\right\}^T,\qquad \mathbf{L}^T\cdot\mathbf{R}=\sum_{s}\frac{\kappa_{s}\Omega_{s}}{\left(\Omega_{s}-\omega\right)^2},  
\end{equation}
are expressed in terms of the eigenvalue, $\omega$, which may be solved from the following implicit equation:
\begin{equation}\label{eq:appeigen}
\sum_{s}\frac{\kappa_{s}\Omega_{s}}{\Omega_{s}-\omega}=1.
\end{equation}
A particular solution of Eq.~(\ref{eq:appeigen}) is $\omega_{(0)}=0$, since $\sum_{s}\kappa_{s}=1$. With only two sorts of ions
Eq.~(\ref{eq:appeigen}) reduces to a linear one and gives the second eigenvalue: $\omega=\kappa_1\Omega_2+\kappa_2\Omega_1$. With three sorts of ions, the second and third eigenvalues are
solved from an easy-to-derive quadratic equation. 
Otherwise, Eq.~(\ref{eq:appeigen}) may be solved numerically using the Newton--Rapson method. Totally, there are $S$ eigenvalues, $\omega_{(\sigma)}$, with left, $\mathbf{L}^T_{(\sigma)}$, and right, $\mathbf{R}_{(\sigma)}$, eigenvectors being orthogonal: $\mathbf{L}^T_{(\sigma)}\cdot\mathbf{R}_{(\sigma^\prime)}=0$ for $\sigma\ne \sigma^\prime$. Herewith, the index, $(\sigma)$, enumerates eigenmodes. Now, we can decompose the force, initial condition and solution of Eq.~(\ref{eq:appvect}) over the orthogonal system of left and right eqgenvectors: 
\begin{equation}\label{eq:appdecomp}
F\mathbf{K}=F\sum_{\sigma)}\frac{\mathbf{R}_{(\sigma)}}{\mathbf{L}^T_{(\sigma)}\cdot\mathbf{R}_{(\sigma)}},\qquad
\mathbf{Z}=\sum_{(\sigma)}{Z_{(\sigma)}\mathbf{R}_{(\sigma)}},\qquad \mathbf{Z}^N=\sum_s{Z^N_{(s)}\mathbf{R}_{(s)}},\qquad Z^N_{(\sigma)}=\frac{\mathbf{L}^T_{(\sigma)}\cdot\mathbf{Z}^N}{\mathbf{L}^T_{(\sigma)}\cdot\mathbf{R}_{(\sigma)}},
\end{equation}
$Z_{(s)}$ being the time-dependent \textit{complex amplitude} for $s$-th eigenmode, obeying the 
equation as follows:
\begin{equation}\label{eq:appampl}
 \frac{d\,Z_{(\sigma)}}{dt}=\frac{F}{\mathbf{L}^T_{(\sigma)}\cdot\mathbf{R}_{(\sigma)}}-j\omega_{(\sigma)} Z_{(\sigma)},
\end{equation}
and $Z^N_{(\sigma)}$ is the amplitude at $t=t^N$. Particularly, for the eigenmode with $\omega_{(0)}=0$ Eq.~(\ref{eq:appampl}) gives:
$$
Z^{N+1}_{(0)}=
\frac{\sum_{s^\prime}{(\rho\mathbf{u})^N_{s^\prime}} +
\Delta t\,F}
{\rho/(\tilde{\rho}B)},
\qquad 
(\rho\mathbf{u})^{N+1}_s\leftarrow(\rho\mathbf{u})^{N+1}_s+\frac{\rho_s}\rho
\left\{\sum_{s^\prime}{(\rho\mathbf{u})^N_{s^\prime}} +
\Delta t\left[]-\frac{\sum_{i^\prime}\left(\Pi_{i^\prime}\cdot\mathbf{S}_{i^\prime}\right)^{N+1/2}}V\right]\right\},
$$
since $\kappa_s/\Omega_s=\rho_s/(\tilde{\rho}B)$. In this eigenmode all ions have a common velocity equal to the total momentum-to-mass  ratio. For other eigenmodes Eq.~(\ref{eq:appampl}) may be solved using the Crtank-Nicolson scheme:
\begin{equation*}
\frac{Z^{N+1}_{(\sigma)}-Z^{N}_{(\sigma)}}{\Delta t}=\frac{F}{\mathbf{L}^T_{(\sigma)}\cdot\mathbf{R}_{(\sigma)}}-j\frac{\omega_{(\sigma)}}2\left[Z^{N+1}_{(\sigma)}+  Z^{N}_{(\sigma)}\right].
\end{equation*}
The latter cam be explicit solved as a linear equation with respect to complex $Z^{N+1}_{(\sigma)}$:
\begin{equation}\label{eq:appmode}
Z^{N+1}_{(\sigma)}=Z^{N}_{(\sigma)}+\frac{F\Delta t}{\mathbf{L}^T_{(\sigma)}\cdot\mathbf{R}_{(\sigma)}}-j\omega_{(\sigma)}\Delta t\,Z^{N+1/2}_{(\sigma)},\qquad Z^{N+1/2}_{(\sigma)}=\left[Z^{N}_{(\sigma)}+\frac{F\Delta t/2}{\mathbf{L}^T_{(\sigma)}\cdot\mathbf{R}_{(\sigma)}}\right]\frac{1-j{\omega_{(\sigma)}\Delta t}/2}{1+{\left(\omega_{(\sigma)}\Delta t\right)^2}/4}.
\end{equation}
The second of Eqs.~(\ref{eq:appmode}) also implicitly advances the solution  of Eq.~(\ref{eq:appampl}) through the time interval equal to $\Delta t/2$ and may be used for this purpose within the multi-ion MHD.  To transform complex amplitudes in Eq.~(\ref{eq:appmode}) back to vectors, one needs to substitute a vector product by $\mathbf{b}$ for each $-j$. Therefore, a stationary solution of Eqs.~(\ref{eq:appmode}), $Z^{N+1}_{(\sigma)}=Z^{N+1/2}_{(\sigma)}=Z^N_{(\sigma)}=-jF/\left(\omega_{(\sigma)}\mathbf{L}^T_{(\sigma)}\cdot\mathbf{R}_{(\sigma)}\right)$, describes the ion drift along the direction of $\mathbf{f}\times\mathbf{b}$. Using Eqs.~(\ref{eq:appeigvec}),(\ref{eq:appeigen}),(\ref{eq:appdecomp}), one can find that only zeroth eigenmode provides non-zero contribution, $\rho Z_{(0)}/\left(\tilde{\rho}B\right)$, to the momentum density, however, all eigenmodes contribute to the current, which effect for non-zeroth eigenmode is due to the said drift:
\begin{equation}\label{eq:appJ}
\mathbf{J}\leftrightarrow \frac1B\sum_{(\sigma)}Z_{(\sigma)}.
\end{equation}

\subsection{Improved CAM Algorithm for the Hybrid Scheme}
Here, we employ the obtained results to improve CAM step in the hybrid algorithm. The inputs are: $\rho^{N+1/2}_s$, $\mathbf{B}^{N+1/2}$, $\mathbf{f}$, and the mixed moments (see Eq.~\ref{eq:rhoustar}),
$(\rho\mathbf{u})^*_s$, for each ion sort. We start by advancing Eqs.~(\ref{eq:appgov}) through a half time-step, accounting for only the effect of force, $\mathbf{f}$, for each sort of ions: $$\left(\rho\hat{\mathbf{u}}\right)^{N+1/2}_s=
\left(\rho\mathbf{u}\right)^*_s+
\frac{\Delta t}2\kappa_s\mathbf{f}.$$ 
Then, the
parallel motion is solved: $$\left(\rho\mathbf{u}\right)^{N+1/2}_s\leftarrow\mathbf{b}\left[\mathbf{b}\cdot\left(\rho\hat{\mathbf{u}}\right)^{N+1/2}_s\right],\qquad
\mathbf{J}^{N+1/2}\leftarrow\sum_s\frac{q_s}{m_s}\left(\rho\mathbf{u}\right)^{N+1/2}_s,$$ 
and only perpendicular components are kept in the predicted velocities, which will then be used to advance velocities via the eigenmode scheme: $$\left(\rho\hat{\mathbf{u}}\right)^{N+1/2}_s\leftarrow \left(\rho\hat{\mathbf{u}}\right)^{N+1/2}_s - \mathbf{b}\left[\mathbf{b}\cdot\left(\rho\hat{\mathbf{u}}\right)^{N+1/2}_s\right].$$ 
Then, for each eigenmode, upon calculating the eigenvalue, $\omega_{(\sigma)}$ and left, $\mathbf{L}^T_{(\sigma)}$, and right, $\mathbf{R}_{(\sigma)}$, eigenvectors, we do the following:
\begin{itemize}
\item{Predicted vector amplitude is calculated:
$$
(\rho\hat{\mathbf{u}})^{N+1/2}_{(\sigma)}=\frac{\mathbf{L}^T_{(\sigma)}\cdot\left\{(\rho\hat{\mathbf{u}})^{N+1/2}_{s}\right\}}{\mathbf{L}^T_{(\sigma)}\cdot\mathbf{R}_{(\sigma)}}.
$$}
\item{For zeroth eigenmode, $(\rho\mathbf{u})^{N+1/2}_{(0)}=(\rho\hat{\mathbf{u}})^{N+1/2}_{(0)}$, otherwise the amplitude is solved from Eq.~(\ref{eq:appmode}):
$$
(\rho\mathbf{u})^{N+1/2}_{(\sigma)}=
\frac{(\rho\hat{\mathbf{u}})^{N+1/2}_{(\sigma)}+\frac{\omega_{(\sigma)}\Delta t}2(\rho\hat{\mathbf{u}})^{N+1/2}_{(\sigma)}\times\mathbf{b}}{1+{\left(\omega_{(\sigma)}\Delta t\right)^2}/4}.
$$}
\item{
The contributions to the ion velocities (if desired!) and to the current density are added:
$$
\left\{\left(\rho\mathbf{u}\right)^{N+1/2}_s\right\}\leftarrow\left\{\left(\rho\mathbf{u}\right)^{N+1/2}_s\right\}+(\rho\mathbf{u})^{N+1/2}_{(\sigma)}\otimes\mathbf{R}_{(\sigma)},\qquad \mathbf{J}^{N+1/2}\leftarrow \mathbf{J}^{N+1/2}+\frac1B(\rho\mathbf{u})^{N+1/2}_{(\sigma)}
$$}
\end{itemize}
Instead of Eqs.~(\ref{eq:CAM}), (\ref{eq:ECAM}), the field (\ref{eq:OhmsLawDiscrCons}) is calculated: $\mathbf{E}^{N+1/2}=\left(\mathbf{f}-\mathbf{J}^{N+1/2}\times\mathbf{B}^{N+1/2}\right)/\tilde{\rho}^{N+1/2}$.
\newcommand{\newblock}{}
\bibliographystyle{yahapj}
\bibliography{references}

\end{document}